\documentclass[conference]{IEEEtran}
\IEEEoverridecommandlockouts
\usepackage{cite}
\usepackage{amsmath,amssymb,amsfonts}
\usepackage{algorithmic}
\usepackage{graphicx}
\usepackage{textcomp}
\usepackage{xcolor}

\usepackage{soul, xcolor}

\usepackage{algorithm}
\usepackage{algorithmic}
\usepackage{setspace}
\usepackage{url}

\def\BibTeX{{\rm B\kern-.05em{\sc i\kern-.025em b}\kern-.08em
    T\kern-.1667em\lower.7ex\hbox{E}\kern-.125emX}}

\makeatletter
\newcommand{\linebreakand}{%
  \end{@IEEEauthorhalign}
  \hfill\mbox{}\par
  \mbox{}\hfill\begin{@IEEEauthorhalign}
}
\makeatother

\begin{document}

\title{Parallelized Multi-Agent Bayesian Optimization in Lava}

\author{\IEEEauthorblockN{1\textsuperscript{st} Shay Snyder}
\IEEEauthorblockA{\textit{Electrical and Computer Engineering} \\
\textit{George Mason University}\\
Fairfax, USA \\
ssnyde9@gmu.edu}

\and
\IEEEauthorblockN{2\textsuperscript{nd} Derek Gobin}
\IEEEauthorblockA{\textit{Electrical and Computer Engineering} \\
\textit{George Mason University}\\
Fairfax, USA \\
dgobin@gmu.edu}

\and
\IEEEauthorblockN{3\textsuperscript{rd} Victoria Clerico}
\IEEEauthorblockA{\textit{Electrical and Computer Engineering} \\
\textit{George Mason University}\\
Fairfax, USA \\
mclerico@gmu.edu}

\linebreakand
\IEEEauthorblockN{4\textsuperscript{th} Sumedh R. Risbud}
\IEEEauthorblockA{\textit{Neuromorphic Computing Lab} \\
\textit{Intel Labs}\\
Santa Clara, USA \\
sumedh.risbud@intel.com}

\and
\IEEEauthorblockN{5\textsuperscript{th} Maryam Parsa}
\IEEEauthorblockA{\textit{Electrical and Computer Engineering} \\
\textit{George Mason University}\\
Fairfax, USA \\
mparsa@gmu.edu}
}

\maketitle

\begin{abstract}

In parallel with the continuously increasing parameter space dimensionality, search and optimization algorithms should support distributed parameter evaluations to reduce cumulative runtime.
Intel's neuromorphic optimization library, Lava-Optimization, was introduced as an abstract optimization system compatible with neuromorphic systems developed in the broader Lava software framework.
In this work, we introduce Lava Multi-Agent Optimization (LMAO) with native support for distributed parameter evaluations communicating with a central Bayesian optimization system.
LMAO provides an abstract framework for deploying distributed optimization and search algorithms within the Lava software framework. 
Moreover, LMAO introduces support for random and grid search along with process connections across multiple levels of mathematical precision.
We evaluate the algorithmic performance of LMAO with a traditional non-convex optimization problem, a fixed-precision transductive spiking graph neural network for citation graph classification, and a neuromorphic satellite scheduling problem.
Our results highlight LMAO's efficient scaling to multiple processes, reducing cumulative runtime and minimizing the likelihood of converging to local optima. 
    

\end{abstract}

\section{Introduction}
Many of today's most interesting problems require solutions to high dimensional and non-linear systems that determine the optimal parameter configuration. Multiple areas such as autonomous robotics~\cite{patton2021neuromorphic}, graph neural networks~\cite{cong2022semi}, evolutionary algorithms~\cite{parsa2021accurate}, and physics-informed neural networks~\cite{scharzenberger2021learning} are limited by the time expenditure from individual parameter evaluations. Rather than traditional procedural approaches like random search~\cite{bergstra2012random} or grid search~\cite{liashchynskyi2019grid}, modern techniques employ heuristic algorithms making informed decisions from prior knowledge, such as Bayesian optimization (BO)~\cite{rasmussen2006gaussian} with roots in Bayesian statistics~\cite{bayes1763lii}. While BO reduces the quantity of problem evaluations by orders of magnitude, many problems still face runtime issues where the reduced number of synchronous evaluations is not enough to compensate for the immense time required by individual evaluations~\cite{cong2023hyperparameter}.

Intel's neuromorphic software framework, Lava, was introduced as an abstract software framework for developing neuromorphic systems.
In this work, we introduce \textbf{L}ava \textbf{M}ulti-\textbf{A}gent \textbf{O}ptimization (LMAO), a novel framework for evaluating parameter configurations across multiple asynchronous processes whose results are aggregated into a single optimizer or search algorithm.
This framework is completely open-sourced through GitHub\footnote{Code available at \url{https://github.com/Parsa-Research-Laboratory/lmao}.}.
We evaluate the performance improvements and operational characteristics of LMAO with the Ackley function~\cite{ackley1987model}, a fixed-precision transductive spiking neural network for citation graph classification~\cite{snyder2024transductive}, and a satellite scheduling problem using quadratic unconstrained binary optimization~\cite{lava}.

In summary, the major contributions of this paper are:
\begin{itemize}
    \item We introduce Lava Multi-Agent Optimization (LMAO) with support for distributed optimization.
    \item We demonstrate the performance of LMAO with the Ackley function~\cite{ackley1987model}, a transductive spiking graph neural network~\cite{snyder2024transductive}, and a QUBO optimization problem for satellite scheduling~\cite{lava}.
\end{itemize}
\section{Architecture of LMAO within the Lava software framework}

Intel's neuromorphic software framework, Lava~\cite{lava}, provides an abstract interface for building interconnected systems of event-based computational elements.
The lowest-level building blocks are \textit{Processes} which provide a blueprint of inputs, outputs, and internal variables.
Lava provides a base library of ports allowing inter-process communication.
\textit{In-Ports} receive information from other processes whereas \textit{Out-Ports} transmit information to other processes.
Individual \textit{Process} functionality is defined within \textit{Process Models}\footnote{See \url{http://lava-nc.org} for details about Lava concepts like \textit{Processes} and \textit{Process Models}}.
Moreover, \textit{Process Models} are architecture specific so the same \textit{Process} can have multiple \textit{Process Models} for execution on different hardware platforms such as central processing units or Loihi 2 neurocores.


\textbf{L}ava \textbf{M}ulti-\textbf{A}gent \textbf{O}ptimization (LMAO) introduces the general-purpose \textit{Solver}. Serving as the single point of entry into the LMAO framework, the \textit{Solver} is a contract between users and developers. This utility provides an abstract interface where users define various parameters such as number of iterations, number of initial points, parameter search spaces, and optimization algorithm types.

\begin{figure}
    \centering
    \includegraphics[width=\linewidth]{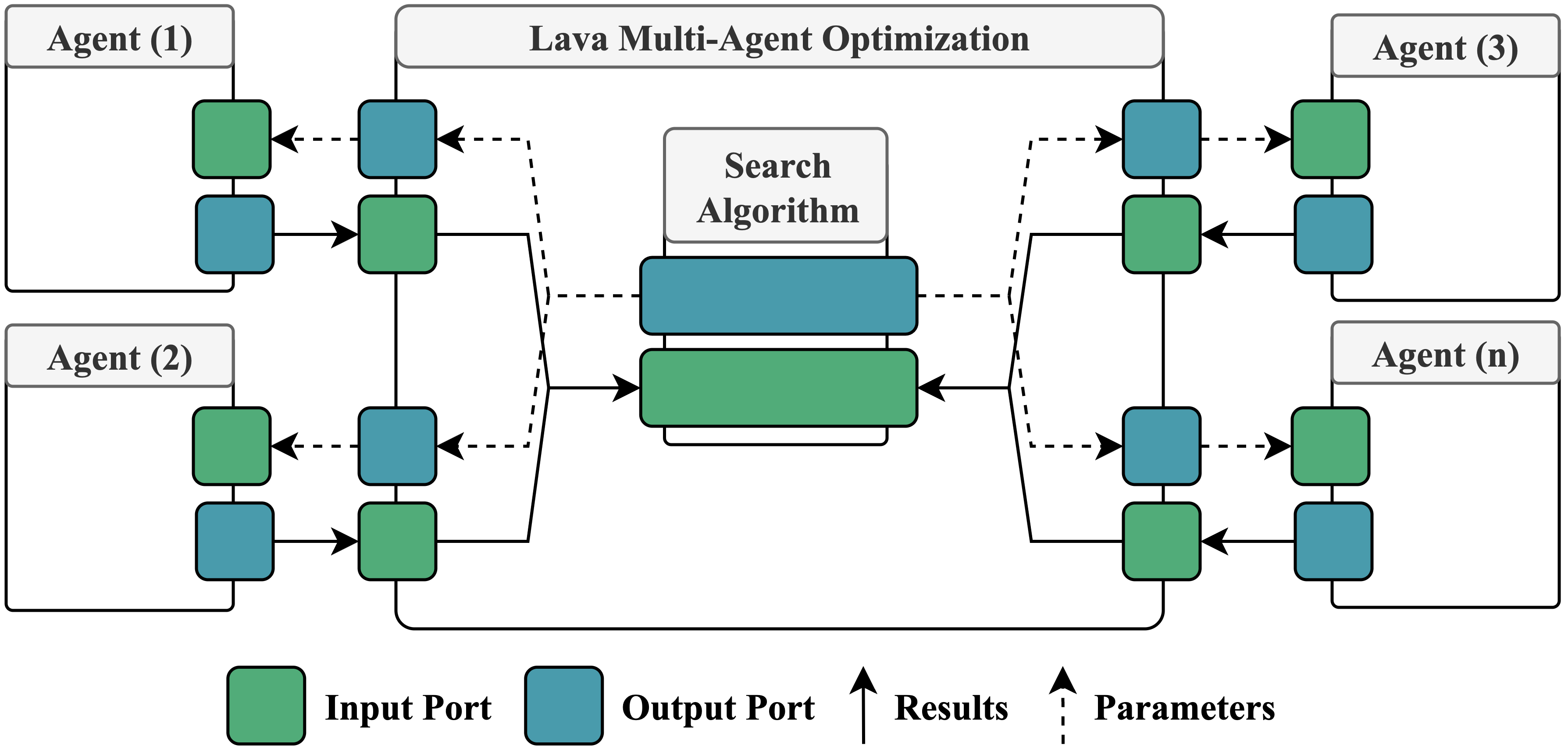}
    \caption{Multiple independently operating agents processing different hyperparameters from the central search algorithm with LMAO.}
    \label{fig:arch-multi-agent}
\end{figure}

\begin{figure*}
    \centering
    \includegraphics[width=\textwidth]{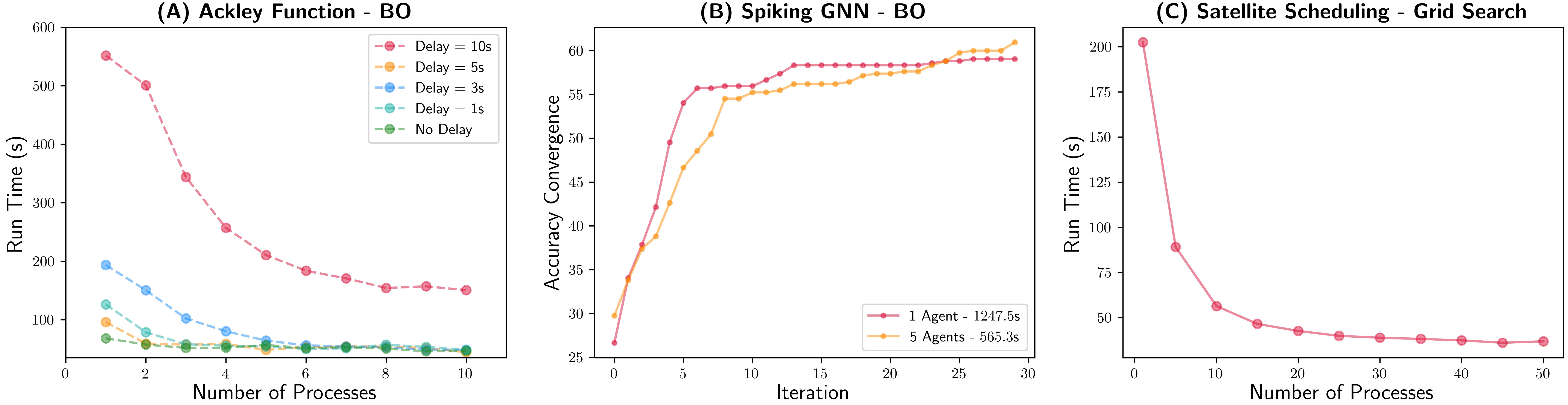}
    \caption[]{(A) The runtime latency of LMAO using BO on the Ackley function~\cite{ackley1987model} with varying amounts of manually induced delay. (B) The accuracy convergence of single and multi-agent BO for citation graph classification~\cite{snyder2024transductive}. (C) Grid search execution times with satellite scheduling~\cite{lava} across different numbers of processes.}
    \label{fig:results}
\end{figure*}

Rather than being limited to sequential parameter evaluations~\cite{snyder2023neuromorphic}, LMAO introduces support for multiple, parallel agents communicating with a central optimization or search algorithm. A high-level flowchart of this process is presented in Figure~\ref{fig:arch-multi-agent}.
Controlled by the \textit{numAgents} parameter, users can distribute evaluations to multiple asynchronous agents and increase the effective number of evaluations per time step.
The LMAO backend supports this functionality by dynamically creating pairs of \textit{In-Ports} and \textit{Out-Ports} for each process and using the Lava runtime framework to distribute agents across multiple processes.

\begin{algorithm}
    \setstretch{1.05}
    \caption{Agent Initialization \& Initial Point Sampling}\label{alg:agent-init}
    \begin{algorithmic}
        \REQUIRE $\mathrm{numIps} = \{\mathrm{numIps} \in \mathbb{N} | \mathrm{numIps} \geq 1\}$
        \REQUIRE numIps $\geq$ numIterations
        \REQUIRE $\mathrm{numAgents} \in \mathbb{N}$
        \REQUIRE $\mathrm{numAgents} \geq 1$
        \STATE opt $\gets$ getOptimizer()
        \STATE ipQueue $\gets$ opt.getInitialPoints(numIps)
        \STATE completedIters $\gets$ 0 
        \FOR{$i=0$ \TO numAgents - 1}
            \item outPort $\gets$ getOutPort(i)
            \item nextPoint $\gets$ ipQueue.pop()
            \item outPort.send(nextPoint)
        \ENDFOR

        \REPEAT  
            \FOR{$i=0$ \TO  $numAgents - 1$}
                \item $\mathrm{inPort} \gets \mathrm{getInPort(i)}$
                \IF {inPort.probe() is $\FALSE$}
                    \item continue
                \ENDIF
                \item $\mathrm{point} \gets \mathrm{inPort.recv()}$
                \item $\mathrm{opt.update(point)}$
                \item $\mathrm{completedIters} = \mathrm{completedIters} + 1$

                \IF{$\mathrm{ipQueue.nonEmpty()}$}
                    \item $\mathrm{outPort} \gets \mathrm{getOutport(i)}$
                    \item nextPoint $\gets$ ipQueue.pop()
                    \item outPort.send(nextPoint)
                \ENDIF

            \ENDFOR
        \UNTIL{$\mathrm{completedIters} \geq \mathrm{numIps}$}
    \end{algorithmic}
\end{algorithm}

As shown in Algorithm~\ref{alg:agent-init}, the system is initialized by sending a unique initial point to each agent. With agents executing asynchronously~\cite{snyder2024asynchronous}, they evaluate the received parameters and return the corresponding values on stochastic time intervals. Simultaneously, the search algorithm probes each \textit{In-Port} from each agent process. If the port has received an evaluated parameter combination, the data is decoded and used to update the search algorithm. This process is repeated until all initial points have been evaluated.

\begin{algorithm}
    \setstretch{1.05}
    \caption{Heuristic Search}\label{alg:heuristic-search}
    \begin{algorithmic}
        \REQUIRE numIps $= \{\mathrm{numIps} \in \mathbb{N} | \mathrm{numIps} \geq 1\}$
        \REQUIRE numIter $ = \{\mathrm{numIter} \in \mathbb{N} | \mathrm{numIter} > \mathrm{numIps}\}$
        \REQUIRE $\mathrm{numAgents} \in \mathbb{N}$
        \REQUIRE $\mathrm{numAgents} \geq 1$
        \STATE opt $\gets$ getOptimizer()
        \STATE completedIters $\gets$ numIps

        \REPEAT
            \STATE numPoints $\gets$ min(numIter - completedIters, numAgents)

            \IF{numPoints $\leq$ 0}
                \RETURN
            \ENDIF
            
            \STATE unknownPoints $\gets$ opt.ask(numPoints)
            \FOR{$i=0$ \TO numPoints - 1}
                \item outPort $\gets$ getOutPort(i)
                \item nextPoint $\gets$ unknownPoints.pop()
                \item outPort.send(nextPoint)
            \ENDFOR
        
            \item numComplete $\gets$ 0
            \REPEAT
                \item $\mathrm{inPort} \gets \mathrm{getInPort(i)}$
                \IF {inPort.probe() is $\TRUE$}
                    \item $\mathrm{point} \gets \mathrm{inPort.recv()}$
                    \item $\mathrm{opt.update(point)}$
                    \item $\mathrm{completedIters} = \mathrm{completedIters} + 1$
                \ENDIF
            \UNTIL{numComplete $\geq$ numPoints}
        \UNTIL{$\mathrm{completedIters} \geq \mathrm{numIter}$}
    \end{algorithmic}
\end{algorithm}

With all initial points evaluated, LMAO uses learned knowledge to heuristically explore the parameter space. As shown in Algorithm~\ref{alg:heuristic-search} and using the constant liar strategy~\cite{10.1007/978-3-642-44973-4_7}, unique, unknown points are sampled and transmitted to each agent for parallelized evaluation. Upon completion, the evaluated points are used to update the model. This process continues until the desired number of iterations is reached wherein all processes are stopped and results are returned to the user.

\section{Results \& Discussion}

We evaluate the performance of Lava Multi-Agent Optimization with a traditional non-convex optimization problem~\cite{ackley1987model}, a spiking graph neural network for citation graph classification~\cite{snyder2024transductive}, and a quadratic unconstrained binary optimization problem in the Lava-Optimization library~\cite{lava}. All experiments were conducted with a desktop computer equipped with an AMD Ryzen 7 3700x processor and 64GB of quad-channel DDR4 memory.

\subsection{\textbf{Traditional Non-Convex Optimization}}

The Ackley~\cite{ackley1987model} function is a classic non-convex optimization problem with widespread usage. We evaluate this function with Bayesian Optimization (BO) across a varying quantity of agents. The search space is continuous, real values across the range of each problem dimension. For a total of 50 optimization iterations, we configure the optimizer to sample 10 initial points before using BO to intelligently explore based on prior knowledge. BO is able to successfully learn each function and converge arbitrarily close to the global minima. As shown in Figure~\ref{fig:results}A, we perform the same experiment across a varying quantity of agents from 1 to 10. Given the low computational complexity of individual function evaluations, the overhead of spawning multiple processes and the latency of updating the Gaussian model outweigh the benefits of multiple agents and doesn't reduce the overall runtime. To evaluate the necessary evaluation latency for LMAO's multi-agent capabilities to be effective, we manually add delay to each function evaluation. As shown in Figure~\ref{fig:results}A, we iterate over multiple delay values: 1s, 3s, 5s, and 10s. These results highlight the positive correlation between the performance benefits from multiple agents and the latency of individual functional evaluations.

\subsection{\textbf{Transductive Spiking Graph Neural Networks}}



In the second experiment, we demonstrate the performance of LMAO with a fixed-precision spiking graph neural network. Introduced in~\cite{cong2022semi}, citation graph classification is performed with transductive learning where the spiking neural network structure is designed based on the citation graph itself. More recent works such as~\cite{cong2023hyperparameter} and~\cite{snyder2024transductive} demonstrate the capability of this approach with Bayesian optimization (BO) while intra-network computations are limited to integer precision compatible with Loihi~\cite{davies2018loihi}.

\begin{table}[]
\caption{The parameter search space for optimizing the fixed-point spiking graph neural network~\cite{snyder2024transductive} with LMAO.}
    \label{tab:gnn-search-space}
    \begin{center}
    \begin{tabular}{c|c}
    \hline
    \textit{\textbf{Parameter}} & \textit{\textbf{Options}}      \\ \hline
    Paper to Paper Weight              &  \{100, 101, ..., 500\} \\  
    Train to Topic Weight              &  \{1, 2, ..., 10\}      \\
    Val. to Topic $\tau_+$ \& $\tau_-$ &  \{20, 21, ..., 60\}    \\
    Simulation Steps                   &  \{5, 7, ..., 13\}      \\ \hline
    \end{tabular}
    \end{center}
\end{table}

Using LMAO, our goal is to reduce the total optimization time required by BO to select the optimal parameter set. As shown in Table~\ref{tab:gnn-search-space}, our search space consists of 4 variables: paper to paper weight, train to topic weight, val to topic $\tau_{+/-}$ and number of simulation steps. We perform two BO experiments with 1 and 5 agents, with the results being averaged over 3 repetitions and 3 random seeds. Both experiments generate and evaluate 10 random points to initialize the underlying Gaussian process (GP). For the experiment with one agent, the acquisition function is used to select a point to evaluate next. This point is evaluated with the results incorporated into the GP. This process is repeated 20 times for 30 total iterations. Conversely, the experiment with 5 agents takes the initialized model and selects 5 unknown points using the constant liar strategy~\cite{10.1007/978-3-642-44973-4_7}. These 5 points are evaluated in parallel with the results returned and used to update the GP. Distributing the optimization across multiple agents reduces the total number of GP model updates and expands the variety of evaluated points. As shown in Figure~\ref{fig:results}B, this allows the experiment with 5 agents to explore a wider area of the search space and avoid converging to local optima as in the case with 1 agent. Moreover, expanding the search across multiple agents reduces the overall optimization time by 2.2x.


\subsection{\textbf{Satellite Scheduling with Quadratic Unconstrained Binary Optimization}}
\begin{table}[]
\caption{The parameter search space for optimizing the neuromorphic satellite scheduling problem~\cite{lava} with LMAO. The overall space contains 270 parameter combinations.}
    \label{tab:satellite-search-space}
    \begin{center}
    \begin{tabular}{c|c}
    \hline
    \textit{\textbf{Parameter}} & \textit{\textbf{Options}} \\ \hline
    Turning Rate             &  \{1.0, 1.25, ..., 3.0\}   \\
    View Height              &  \{0.25, 0.50, ..., 1.5\}  \\
    Number of Satellites     &  \{2, 3, 4, 5, 6\} \\ \hline
    \end{tabular}
    \end{center}
\end{table}

In our last experiment, we highlight the performance impact of multi-agent optimization for traditional grid search applied to a novel satellite scheduling algorithm within the Lava-Optimization library~\cite{lava}. Using the 270 parameter search space shown in Table~\ref{tab:satellite-search-space}, we perform grid search with varying numbers of agents, ranging from 1 to 50. As shown in Figure~\ref{fig:results}C, LMAO's multi-agent architecture efficiently scales where there is an inverse correlation between the number of agents and total search time. Specifically, increasing the number of agents from 1 to 50 reduced cumulative runtime by 5.57x.

\section{Conclusion}
In this work, we introduce \textbf{L}ava \textbf{M}ulti-\textbf{A}gent \textbf{O}ptimization (LMAO), a novel framework for parallelized optimization and search algorithms within the Lava software framework. Our results demonstrate the scalability of this system applied to a variety of application spaces with multiple optimization and search algorithms.
Using the abstract framework provided by LMAO, we are planning to include more algorithms such as: evolutionary algorithms~\cite{De_Jong2016-tu}, hyperdimensional Gaussian process regression~\cite{furlong2022fractional} and distributed Bayesian search~\cite{hyperspace}.
Moreover, we will expand the application space for LMAO in areas such as automated neural network design~\cite{parsa2021accurate} and robotic control~\cite{schuman2022evolutionary}.
\section{Acknowledgements}
The research was funded in part by National Science Foundation
through award CCF2319619 and a gift from Intel Corporation.

\bibliographystyle{IEEEtran}
\bibliography{sample-base}

\end{document}